\documentclass{article}

\usepackage{amssymb,amsfonts,amsmath,stmaryrd}
\usepackage{cite,enumerate,float,indentfirst}
\usepackage{color}
\usepackage{tikz}
\usetikzlibrary{arrows,snakes,backgrounds}
\usepackage{hyperref}

\def\be{\begin{eqnarray}}
\def\ee{\end{eqnarray}}
\def\nn{\nonumber}

\def\p{\partial}
\def\tr{{\rm tr}\,}
\def\Tr{{\rm Tr}\,}

\def\b{{\rm b}}

\definecolor{red}{rgb}{1,0,0}
\definecolor{orange}{rgb}{1,0.5,0}
\definecolor{violet}{rgb}{0.7,0,1}

\hoffset= - 1.0in         

\begin{document}

\begin{center}
\begin{small}
\hfill FIAN/TD-12/21\\
\hfill IITP/TH-19/21\\
\hfill ITEP/TH-29/21\\
\hfill MIPT/TH-16/21\\
\end{small}
\end{center}

\vspace{.5cm}

\begin{center}
\begin{Large}\fontfamily{cmss}
\fontsize{15pt}{27pt}
\selectfont
	\textbf{From superintegrability to tridiagonal representation of $\beta$-ensembles}
	\end{Large}
	
\bigskip \bigskip

\begin{large}A. Mironov$^{a,b,c,}$\footnote{mironov@lpi.ru; mironov@itep.ru},
A. Morozov$^{d,b,c,}$\footnote{morozov@itep.ru},
A. Popolitov$^{d,b,c,}$\footnote{popolit@gmail.com}
 \end{large}
\\
\bigskip

\begin{small}
$^a$ {\it Lebedev Physics Institute, Moscow 119991, Russia}\\
$^b$ {\it ITEP, Moscow 117218, Russia}\\
$^c$ {\it Institute for Information Transmission Problems, Moscow 127994, Russia}\\
$^d$ {\it MIPT, Dolgoprudny, 141701, Russia}
\end{small}
 \end{center}

\bigskip

\begin{abstract}
  The wonderful formulas by I.Dumitriu and A.Edelman \cite{paper:DE-matrix-models-for-beta-ensembles,
    paper:K-on-gaussian-random-matrices-coupled-to-the-discrete-laplacian}
rewrite $\beta$-ensemble, with eigenvalue integrals containing Vandermonde factors in the power $2\beta$, through integrals over tridiagonal matrices, where $\beta$-dependent are the powers of individual matrix elements, not their differences. These potentially useful formulas are usually deduced from rather complicated and non-transparent combinatorics and are not as widely known as they deserve. We explain that the superintegrability property, i.e. a simple expression of the Gaussian averages of arbitrary Jack polynomials through the same Jack polynomials, is immediately consistent with this tridiagonal representation, which may serve as a clue to its simple and transparent interpretation. For a formal non-perturbative proof, we use the Virasoro constraints, which themselves acquire an interesting structure in the tridiagonal realization. We also attract attention to the surprising spontaneous breakdown of discrete invariance by the tridiagonal measure, which may signal a new interesting anomaly at the elementary level of the basic eigenvalue matrix model.
\end{abstract}

\bigskip

\section{Introduction}

Eigenvalue matrix models
\cite{
  book:M-random-matrices,
  paper:IZ-the-planar-approximation-II,
  Marshakov:1991gc,
  Kharchev:1992iv,
  paper:M-integrability-and-matrix-models,
  paper:M-2dgravity,
    paper:M-matrix-models-as-integrable-systems
}
are integrals over Hermitian matrices $M$ of expressions
which depend only on the time-variables $p_k=\Tr M^k= \sum_{i=1}^N \lambda_i^k$,
i.e. on the
eigenvalues $\lambda_i$ of $M=U^\dagger D U$, with diagonal matrix $D = {\rm diag}(\lambda)$.
Then one can integrate over the angular coordinates (unitary matrices) $U$
and get an integral over eigenvalues $\lambda$ only
but with a non-trivial weight $\Delta(\lambda)^2$,
made from the Vandermonde determinant $\Delta(\lambda) = \prod_{i<j}^N(\lambda_i-\lambda_j)$.
One can further deform the power of the Vandermonde determinant in order to introduce $\beta$-ensemble
\be
\Big< F\{p_k\}\Big> :=  \int_{-\infty}^\infty
\Delta(\lambda)^{2\beta}\,  \prod_{i=1}^N  e^{-\lambda_i^2/2} d\lambda_i
\cdot F\{p_k\}
\label{evMAMO}
\ee
where $F\{p_k\}$ are polynomials of $p_k$.

The $\beta$-ensemble has many applications in modern statistical and mathematical physics
(in simplest applications, it is convenient to make it Gaussian as we do, and
all non-Gaussianities are then included into $F$).
In particular, $\beta = 1/2$ and $\beta=2$ cases are associated with integrals over orthogonal
and symplectic matrices $U$, while an arbitrary $\beta$ arises in the theory of DIM algebras
\cite{
  paper:AKMMMOZ-generalized-kz-equation-for-dim,
  paper:MMZ-dim-symmetry-of-network-matrix-models
}
(where $t=q^\beta$), and therefore in theory of the $5d$ AGT relations
\cite{
  paper:MMS-proving-agt-as-hs-duality,
  Itoyama:2016dwe,
  paper:NZ-q-virasoro-constraints-in-matrix-models,
  paper:ZP-localization-review,
  paper:MNZ-5d-sym-and-ads7,
  paper:NNZ-q-virasoro-modular-double-and-3d-partition-functions
}
and Nekrasov functions \cite{
  paper:IO-method-of-generating-q-expansion-coefficients,
    paper:NS-quantization-of-integrable-systems-and-four-dimensional-gauge-theories,
  paper:MM-nekrasov-functions-from-exact-bs-periods-sun,
  paper:MM-nekrasov-functions-and-exact-bs-integrals,
  paper:NRS-darboux-coordinates-yy-functional-and-gauge-theory
}.
Dealing with arbitrary powers of Vandermonde determinant is, however, a technically complicated problem,
which makes the entire field somewhat advanced and difficult.
Therefore any alternative approach can be useful, at least to widen view on the problem.

Among such approaches a very interesting one is an old observation
\cite{
  paper:DE-matrix-models-for-beta-ensembles,
  paper:K-on-gaussian-random-matrices-coupled-to-the-discrete-laplacian
}
that
the $\beta$-deformation looks technically simpler if one reduces the matrix integrals over $M$
only partly: not to diagonal matrices, but to tridiagonal ones,
\be
    M \ \longrightarrow \   \left(\begin{array}{ccccc}
      a_1 & \b_1 & 0 & 0 & \dots \\
      \bar \b_1 & a_2 & \b_2 & 0 & \\
      0 & \bar \b_2 & a_3 & \b_3 & \\
      0 & 0 & \bar \b_3 & a_4 & \\
      \dots
    \end{array} \right)
\ee
Tri-diagonal matrices are distinguished because all $p_k=\tr M^k$ contain all $\b_j$
only in combinations $b_j\bar b_j$, and hence the integrals of the phases of $\b_j$ decouple.
Thus tridiagonal integrals naturally comes in terms of {\it real}-valued positive ``radii" $\b_j$,
i.e. go over matrices
\be\label{Psi}
\Psi = \left(\begin{array}{ccccc}
      a_1 & \b_1 & 0 & 0 & \dots \\
        \b_1 & a_2 & \b_2 & 0 & \\
      0 &   \b_2 & a_3 & \b_3 & \\
      0 & 0 &   \b_3 & a_4 & \\
      \dots
    \end{array} \right)
\ee
with $\b_j\geq 0$.
The remarkable result of \cite{
  paper:DE-matrix-models-for-beta-ensembles,
  paper:K-on-gaussian-random-matrices-coupled-to-the-discrete-laplacian
} is that, in terms of $\Psi$, the average  (\ref{evMAMO}) becomes
\be
\Big< F\{p_k\}\Big> \sim
    \prod_{i=1}^N \int_{-\infty}^\infty e^{-a_i^2/2} d a_i \prod_{j=1}^{N-1}
    \int_{0}^\infty e^{-\b_j^2} \b_j^{2\beta j - 1} d\b_j \cdot F\{\tr \Psi^k\}
\label{triMAMO1}
\ee
i.e. raised to the power $\beta$ are just individual matrix elements $\b_i$
and not their combinations (differences), like it was in the case of eigenvalues $\lambda_i$ in (\ref{evMAMO}).
Note that the permutation symmetry $\lambda_i \rightarrow \lambda_{\sigma(i)}$,
apparent in the eigenvalue representation \eqref{evMAMO}, is hidden in the tridiagonal form
\eqref{triMAMO}, and manifests itself through emergence of zero modes
(see sec.~\ref{sec:non-perturbative-derivation}).

Now one notices that the traces of matrix do not change if one chooses the matrix instead of (\ref{Phi}) in the form
\be
 \left(\begin{array}{ccccc}
      a_1 & \b_1^2 & 0 & 0 & \dots \\
        1 & a_2 & \b_2^2 & 0 & \\
      0 &  1 & a_3 & \b_3^2 & \\
      0 & 0 &  1 & a_4 & \\
      \dots
    \end{array} \right)
\ee
Indeed, this transformation is achieved by rotation, $\Psi\longrightarrow U^{-1}\cdot\Psi\cdot U$
with the diagonal matrix $U$ such that $U_{ij}=\prod_{k=1}^{i-1}\b_k\cdot\delta_{ij}$.
Then one naturally makes the change of variables $b_i=\b_i^2$ in the integral (\ref{triMAMO1}) in order to obtain
\be
\boxed{
\Big< F\{p_k\}\Big> \sim
    \prod_{i=1}^N \int_{-\infty}^\infty e^{-a_i^2/2} d a_i \prod_{j=1}^{N-1}
    \int_{0}^\infty e^{-b_j} b_j^{\beta j - 1} db_j \cdot F\{\tr \Phi^k\}
}
\label{triMAMO}
\ee
with
\be\label{Phi}
\Phi = \left(\begin{array}{ccccc}
      a_1 & b_1  & 0 & 0 & \dots \\
        1 & a_2 & b_2  & 0 & \\
      0 &  1 & a_3 & b_3  & \\
      0 & 0 &  1 & a_4 & \\
      \dots
    \end{array} \right)
\ee

\section{Superintegrability}

In \cite{
  paper:DE-matrix-models-for-beta-ensembles,
  paper:K-on-gaussian-random-matrices-coupled-to-the-discrete-laplacian
} the result (\ref{triMAMO}) was deduced from some complicated combinatorics,
which made it somewhat obscure and non-evident.
The claim of the present paper is that one can simplify the problem by establishing
the equivalence (\ref{evMAMO}) = (\ref{triMAMO}) for a complete set of polynomials $F\{p_k\}$.
A natural choice of the basis in the space of polynomials is provided by the Jack polynomials ($\beta$-deformation of the Schur polynomials), because then (\ref{evMAMO}) possesses the remarkable property of {\it superintegrability}
\cite{
  paper:MM-on-the-complete-perturbative-solution-of-on-matrix-models,
  paper:IMM-tensorial-generalization-of-characters,
    paper:MM-sum-rules-for-characters-from-characters-preservation-property-of-matrix-models
}:
the (normalized) averages of the Jack polynomials are again Jack polynomials \cite{
  paper:MMS-towards-a-proof-of-agt-mm,
paper:MPSh-on-qt-deformation-of-gaussian-mm
},
\be
\left(\prod_{m=1}^{N}{\Gamma(\beta+1)\over (2\pi)^{1/2} \Gamma(m\beta+1)}\right)\cdot  \int_{-\infty}^\infty
 \Delta(\lambda)^{2\beta}
 \prod_{i=1}^N e^{-\lambda_i^2/2} d\lambda_i\cdot
\cdot J_R\left\{p_k=\sum_{i=1}^N\lambda_i^k\right\}
= \beta^{|R|\over 2}\cdot\frac{J_R\{N\}J_R\{\delta_{k,2}\}}{J\{\delta_{k,1}\}}
\label{Jackave}
\ee
Then it remains to claim the same for (\ref{triMAMO}), and this is indeed true:

\underline{for $\beta=1$}
\be
 \frac{1}{(2\pi)^{N/2}\prod_{i=0}^{N-2} i!} \cdot \prod_{i=1}^N\int_{-\infty}^\infty  e^{-a_i^2/2}  da_i
  \prod_{i=1}^{N-1} \int_0^\infty e^{-b_i} b_i^{i-1}db_i \ \   \chi_R\{p_k=\tr \Phi^k\}
  = \frac{\chi_R\{N\}\chi_R\{\delta_{k,2}\}}{\chi_R\{\delta_{k,1}\}}
\ee
for the Schur polynomials $\chi_R\{p_k\}$ labeled by arbitrary Young diagrams $R$.
For generic $\beta$, the Schur polynomials are substituted by the Jack polynomials

\underline{generic $\beta$}
\be\label{J}
\boxed{
\frac{1}{(2\pi)^{N/2}\prod_{m=1}^{N-1} \Gamma (m\beta)}\cdot
\prod_{i=1}^N\int_{-\infty}^\infty \!\!\!\! e^{-a_i^2/2}  da_i
\prod_{i=1}^{N-1} \int_0^\infty \!\!\!\! e^{-b_i} b_i^{\beta i-1}db_i \ \   J_R\{p_k=\tr \Phi^k\}
= \beta^{|R|\over 2}\cdot\frac{J_R\{N\}J_R\{\delta_{k,2}\}}{J_R\{\delta_{k,1}\}}
}
\ee

\section{Consistency check }

The formula for the averages (\ref{J}) is consistent with two elementary sum rules
involving Gaussian integrals.
As usual, this helps to understand why it is natural.
In intermediate formulas in this section, we use
the normalization of the Jack polynomials as in \cite{
paper:MMS-towards-a-proof-of-agt-mm
}.
The normalization does not affect (\ref{Jackave}) and (\ref{J}).

The Cauchy formula states that
\be
\sum_R||J_R||J_R\{p_k\}J_R\{p'_k\}=\exp\left(\sum_{k=1}{\beta p_kp'_k\over k}\right)
\ee
where
\be
||J_R||:=\prod_{i,j}{R_i-j+\beta(R^\vee_j-i+1)\over R^\vee_i-j+\beta^{-1}(R_j-i+1)}
\ee
and $R^\vee$ denotes the transposed Young diagram.

Now, multiplying the l.h.s. of (\ref{J}) by $J_R(\delta_{k,1})||J_R||$ and summing over $R$, one obtains
\be
  \frac{1}{(2\pi)^{N/2}\prod_{m=1}^{N-1} \Gamma (m\beta)}\prod_{i=1}^N\int_{-\infty}^\infty  e^{-a_i^2/2}  da_i
  \prod_{i=1}^{N-1} \int_0^\infty e^{-b_i} b_i^{\beta i-1}db_i \ \   \sum_R||J_R||J_R\{p_k=\tr \Phi^k\}J_R(\delta_{k,1})=\nn\\
=     \frac{1}{(2\pi)^{N/2}\prod_{m=1}^{N-1} \Gamma (m\beta)}\prod_{i=1}^N\int_{-\infty}^\infty  e^{-a_i^2/2}  da_i
  \prod_{i=1}^{N-1} \int_0^\infty e^{-b_i} b_i^{\beta i-1}db_i\exp\Big(\beta\tr\Phi\Big)=\nn\\
=   \frac{1}{(2\pi)^{N/2}\prod_{m=1}^{N-1} \Gamma (m\beta)} \prod_{i=1}^N\int_{-\infty}^\infty  e^{-a_i^2/2+\beta a_i}  da_i
  \prod_{i=1}^{N-1} \int_0^\infty e^{-b_i} b_i^{\beta i-1}db_i=\nn\\
=  \frac{1}{(2\pi)^{N/2}\prod_{m=1}^{N-1} \Gamma (m\beta)} (2\pi)^{N/2}e^{N\beta^2/2} \prod_{i=1}^{N-1}\Gamma(i\beta )
=e^{N\beta^2/2}=\sum_R \beta^{|R|/2}||J_R||J_R\{N\}J_R\{\delta_{k,2}\}
\ee
which is the r.h.s. of (\ref{J}) summed over $R$ with $J_R(\delta_{k,1})||J_R||$.

Similarly, multiplying the l.h.s. of (\ref{J}) by $J_R(\delta_{k,2})||J_R||$ and summing over $R$, one obtains
\be\label{eq:2}
  \frac{1}{(2\pi)^{N/2}\prod_{m=1}^{N-1} \Gamma (m\beta)}\prod_{i=1}^N\int_{-\infty}^\infty  e^{-a_i^2/2}  da_i
  \prod_{i=1}^{N-1} \int_0^\infty e^{-b_i} b_i^{\beta i-1}db_i \ \   \sum_R||J_R||J_R\{p_k=\tr \Phi^k\}J_R(\delta_{k,2})=\nn\\
=     \frac{1}{(2\pi)^{N/2}\prod_{m=1}^{N-1} \Gamma (m\beta)}\prod_{i=1}^N\int_{-\infty}^\infty  e^{-a_i^2/2}  da_i
  \prod_{i=1}^{N-1} \int_0^\infty e^{-b_i} b_i^{\beta i-1}db_i\exp\Big({1\over 2}\beta\tr\Phi^2\Big)=\nn\\
=   \frac{1}{(2\pi)^{N/2}\prod_{m=1}^{N-1} \Gamma (m\beta)} \prod_{i=1}^N\int_{-\infty}^\infty  e^{-a_i^2/2+\beta a_i^2/2}  da_i
  \prod_{i=1}^{N-1} \int_0^\infty e^{-b_i+\beta b_i} b_i^{\beta i-1}db_i=\nn\\
=  \frac{1}{(2\pi)^{N/2}\prod_{m=1}^{N-1} \Gamma (m\beta)} {(2\pi)^{N/2}\over (1-\beta)^{N/2}} \prod_{i=1}^{N-1}(1-\beta)^{-i\beta}\Gamma(i\beta )
={1\over (1-\beta)^{\beta N^2/2+(1-\beta)N/2}}=\nn\\
=\sum_R \beta^{|R|/2}||J_R||{J_R\{N\}J_R\{\delta_{k,2}\}^2\over J_R\{\delta_{k,1}\}}
\ee
which is the r.h.s. of (\ref{J}) summed over $R$ with $J_R(\delta_{k,2})||J_R||$.
The last equality in (\ref{eq:2}) follows from the expansion
\be
{1\over (1-\beta)^a}=\sum_{m=0}{\beta^m\over m!}\prod_{k=0}^{m-1}(a+k)
\ee
and from the combinatorial identity
\be
\sum_{R:\ {R}=2m} \beta^{|R|/2}||J_R||{J_R\{N\}J_R\{\delta_{k,2}\}^2\over J_R\{\delta_{k,1}\}}=
{\beta^m\over m!}\prod_{k=0}^{m-1}\Big(\beta N^2/2+(1-\beta)N/2+k\Big)
\ee

\section{Comment on the derivation of (\ref{J})}

Constructing an elegant and universal way to prove superintegrability relations like (\ref{J})
is an open problem since they have been explicitly formulated in \cite{
    paper:MM-on-the-complete-perturbative-solution-of-on-matrix-models
    }
and derived in \cite{
paper:MM-correlators-in-tensor-models-from-character-calculus
} from the Wick theorem.
The fact that they can serve as an elementary substitute of the lengthy
direct calculation in \cite{
  paper:DE-matrix-models-for-beta-ensembles,
  paper:K-on-gaussian-random-matrices-coupled-to-the-discrete-laplacian
},
can attract new attention to this problem,
especially of experts in statistics, who know how to solve them
in different ways.
Still, at this moment the question remains open,
and we make a choice in favor of a proof of (\ref{triMAMO})
which is standard for string physicists, though somewhat transcendental
from the point of view of the ``down-to-Earth" subject like (\ref{J}).

\section{Non-perturbative derivation of tridiagonal formula (\ref{triMAMO})}
\label{sec:non-perturbative-derivation}

Even if it was available, one could prefer not to deal with specific bases of polynomials, and deal instead
with an approach based on Virasoro constraints, where the polynomials are substituted by an exponential
$F\{p_k\} = \exp\left\{\sum_k t_k p_k\right\}$ treated as a formal power series.

The point is that the partition function of the $\beta$-ensemble
\be\label{Zbeta}
Z\{t_k\} = \int \Delta^{2\beta}(\lambda)\,
\prod_{i=1}^N e^{-\lambda_i^2/2+\sum_k t_k \lambda_i^k} d\lambda_i
\ee
satisfies the Virasoro constraints
\be
\hat L_nZ =0, \ \ \ \ n\geq -1,
\ee
with
\be
\hat L_{n} = e^{-N t_0}\left [
- \frac{\partial}{\partial t_{n+2}} + (n+1)(1 - \beta)  \frac{\partial}{\partial t_{n}}
 +\sum_{k=1}^\infty k   t_k
\frac{\partial}{\partial t_{k + n}}
+ \beta \sum_{a = 0}^{n} \frac{\partial^2}{\partial t_a \partial t_{n-a}}
\right ] e^{N t_0}
\ee
Moreover, these constraints unambiguously fix the solution (\ref{Zbeta}),
\cite{
  paper:David,
  paper:MM-on-the-origin-of-virasoro-constraints,
  paper:AM-properties-of-loop-equations,
  paper:IM-noncricital-virasoro-algebra
}.

All these Virasoro constraints are generated by the two constraints: the $\beta$-independent constraint
\be
\hat L_{-1} =   Nt_1 - \frac{\partial}{\partial t_1}
+ \sum_{k=2}^\infty kt_k\frac{\partial}{\partial t_{k-1}}
\ee
and the $\beta$-dependent one
\be
\hat L_2 = \Big(2 \beta N+3(1-\beta)\Big) \frac{\partial}{\partial t_2} - \frac{\partial}{\partial t_4}
+ \beta \frac{\partial^2}{\partial t_1^2}
+ \sum_{k=1}^\infty kt_k\frac{\partial}{\partial t_{k+2}}
\ee

Hence, it is sufficient to prove that these two operators annihilate the
tridiagonal version of the partition function,
\be
{\cal Z}\{t_k\} =  \prod_{i=1}^N  \int
e^{-a_i^2/2} da_i \prod_{j=1}^{N-1} \int_0^\infty e^{-b_j} b_j^{\beta j-1} db_j
\exp\left\{\sum_k t_k \,\tr \Phi^k\right\}
\ee

For illustrative purposes, we present calculations for the first four Virasoro constraints, which are of growing complication.

\begin{itemize}
\item[\fbox{$L_{-1}$}] Applying $\hat L_{-1}$ operator to the full partition function, one gets
\be
\hat L_{-1} {\cal Z} =
\prod_{i=1}^N  \int
e^{-a_i^2/2} da_i \prod_{j=1}^{N-1} \int_0^\infty e^{-b_j} b_j^{\beta j-1} db_j
\left(Nt_1 -  \tr \Phi + \sum_{k=2} kt_k \,\tr \Phi^{k-1}\right)
\exp\left\{\sum_k t_k \,\tr \Phi^k\right\} = \nn \\
 = \prod_{i=1}^N  \int
e^{-a_i^2/2} da_i \prod_{j=1}^{N-1} \int_0^\infty e^{-b_j} b_j^{\beta j-1} db_j
\left( -  \sum_{i=1}^N a_i + \sum_{i=1}^N \frac{\p}{\p a_i}\right)
\exp\left\{\sum_k t_k \,\tr \Phi^k\right\}
\ \ \ \ \
\ee
which vanishes as a corollary of the simple identity
\be
\sum_{i=1}^N \frac{\p}{\p a_i} \tr \Phi^k = k\, \tr \Phi^{k-1}
\label{rel-1}
\ee
for the matrix (\ref{Phi}):
it just remains to integrate over $a_i$ by parts to get the required
$\hat L_{-1}{\cal Z} = 0$.
Note that this calculation is independent of the form of the $b$-integral.

\item[\fbox{$L_{0}$}] Likewise, from
\be
\left(\sum_{i=1}^N a_i\frac{\p}{\p a_i} + 2\sum_{j=1}^N b_j\frac{\p}{\p b_j}\right)
\tr \Phi^k = k\, \tr \Phi^{k}
\label{rel0}
\ee
we get
\be
\!\!\!\!\!\!\!\!\!\!\!\!\!\!
\hat L_{0} {\cal Z} =
\prod_{i=1}^N  \int
e^{-a_i^2/2} da_i \prod_{j=1}^{N-1} \int_0^\infty e^{-b_j} b_j^{\beta j-1} db_j
\left(N^2\beta+N(1-\beta ) -  \tr \Phi^2 + \sum_{k=1} kt_k \,\tr \Phi^{k}\right)
\exp\left\{\sum_k t_k \,\tr \Phi^k\right\} = \nn
\ee
{\footnotesize
$$
\!\!\!\!\!\!\!\!\!\!\!\!\!\!\!\!\!\!
= \prod_{i=1}^N  \int
e^{-a_i^2/2} da_i \prod_{j=1}^{N-1} \int_0^\infty e^{-b_j} b_j^{\beta j-1} db_j
\left( N^2\beta+N(1-\beta ) -  \sum_{i=1}^N a_i^2 -2\sum_{j=1}^{N-1} b_j
+ \sum_{i=1}^N a_i\frac{\p}{\p a_i}
+ 2\sum_{j=1}^{N-1} b_j\frac{\p}{\p b_j}\right)
\exp\left\{\sum_k t_k \,\tr \Phi^k\right\}
$$
}

\noindent
which is again zero after we integrate by parts, this time over all $a_i$ and $b_j$. \\

\item[\fbox{$L_{1}$}] In this case, we use the identity
\be\label{zm1}
\!\!\!\!\!\!\!\!\!\!\!
\left(\sum_{i=1}^N a_i^2\frac{\p}{\p a_i} +  \sum_{j=1}^{N-1} \Big((3-2j+c)a_j+(2j+1-c)a_{j+1}\Big)b_j\frac{\p}{\p b_j}
+ \ \ \ \ \ \ \ \ \ \ \ \ \ \ \ \ \ \ \ \ \ \ \ \ \ \ \
\right.\nn \\ \left.
+\sum_{j=1}^{N-1}\Big((2j-c)b_j \frac{\p}{\p a_j}
+(c+2-2j)b_j\frac{\p}{\p a_{j+1}} \Big)
\right)
\tr \Phi^k = k\, \tr \Phi^{k+1}
\ee

\noindent
where $c$ is an arbitrary free parameter or a ``zero mode" (equality holds for any value of $c$).
Therefore
{\footnotesize
\be
\hat L_{1} {\cal Z} =
\prod_{i=1}^N  \int
e^{-a_i^2/2} da_i \prod_{j=1}^{N-1} \int_0^\infty e^{-b_j} b_j^{\beta j-1} db_j
\left( \Big(2N\beta+2(1-\beta)\Big)\cdot \tr \Phi  -  \tr \Phi^3 + \sum_{k=1} kt_k \,\tr \Phi^{k+1}\right)
\exp\left\{\sum_k t_k \,\tr \Phi^k\right\} = \nn
\ee
}
{\footnotesize
\be
= \prod_{i=1}^N  \int
e^{-a_i^2/2} da_i \prod_{j=1}^{N-1} \int_0^\infty e^{-b_j} b_j^{\beta j-1} db_j
\left\{ \Big(2N\beta+2(1-\beta)\Big) \sum_{i=1}^N a_i -\sum_{i=1}^N a_i^3
-3\sum_{j=1}^{N-1}(a_j+a_{j+1})b_j  + \sum_{i=1}^N a_i^2\frac{\p}{\p a_i}
+ \right. \nn \\ \left.
+ \sum_{j=1}^{N-1} \Big((3-2j+c)a_j+(2j+1-c)a_{j+1}\Big)b_j\frac{\p}{\p b_j}
+\sum_{j=1}^{N-1}\Big((2j-c)b_j\frac{\p}{\p a_j} +(c+2- 2j)b_j\frac{\p}{\p a_{j+1}}\Big)
\right\}
\exp\left\{\sum_k t_k \,\tr \Phi^k\right\}
\nn
\ee
}

\noindent
Integration by parts gives:
{\footnotesize
\be
 \! \! \!
\hat L_{1}{\cal Z} = \prod_{i=1}^N  \int
e^{-a_i^2/2} da_i \prod_{j=1}^{N-1} \int_0^\infty e^{-b_j} b_j^{\beta j-1} db_j
\left\{ \Big(2N\beta+2(1-\beta)\Big) \sum_{i=1}^N a_i -\sum_{i=1}^N a_i^3
-3\sum_{j=1}^{N-1}(a_j+a_{j+1})b_j + \sum_{i=1}^N (a_i^3-2a_i)\
+ \ \ \ \ \ \ \ \ \ \ \ \right. \nn \\ \left.
+ \sum_{j=1}^{N-1} \Big((3-2j+c)a_j+(2j+1-c)a_{j+1}\Big)(b_j-\beta j)
+\sum_{j=1}^{N-1}\Big((2j-c)a_jb_j  +(c+2- 2j)a_{j+1}b_j \Big)
\right\}
\exp\left\{\sum_k t_k \,\tr \Phi^k\right\}
= \ \ \ \ \ \ \  \ \ \  \ \ \ \ \ \    \nn
\ee
$$
 \! \! \! \! \! \!\! \! \! \! \! \!
=\beta\cdot  \prod_{i=1}^N  \int
e^{-a_i^2/2} da_i \prod_{j=1}^{N-1} \int_0^\infty e^{-b_j} b_j^{\beta j-1} db_j
\left\{ 2(N-1)   \sum_{i=1}^N a_i
-  \sum_{j=1}^{N-1} \Big(j(3-2j+c)a_j+j(2j+1-c)a_{j+1}\Big)
\right\}
\exp\left\{\sum_k t_k \,\tr \Phi^k\right\}
$$
}

\noindent
which can be made zero for the special choice of $c=2N-3$ .
In fact, the integral at the r.h.s. is independent of $c$,
but this particular choice makes cancellation apparent.

\item[\fbox{$L_{2}$}] \textbf{Finally}, we are ready to prove the main $L_{2}$ constraint,
which together with simply looking $L_{-1}$ generates the entire algebra
of Virasoro constraints.
In this case, we need
\be
\left(\sum_{i=1}^N a_i^3\frac{\p}{\p a_i}
+ \sum_{l=1}^{N-1} b_l \sum_{i,j=1}^N u_l^{ij}a_i \frac{\p}{\p a_j}
+\sum_{l=1}^{N-1} b_l \sum_{i,j=1}^N v_l^{ij}a_ia_j \frac{\p}{\p b_l}
+ \sum_{i,j =1}^{N-1} w^{ij}b_ib_j\frac{\p}{\p b_j} \right)
\tr \Phi^k = k\, \tr \Phi^{k+2}
\label{2id}
\ee
where we have a whole variety of zero modes in matrices $u_l$, $v_l$ and $w$.

These zero modes can be chosen in such a way that (see the general expression in \eqref{eq:full-k-2-operator})
\be
u^{ij}_l=2(N-l)\Big(\delta_{i,j}\delta_{j,l+1}+\delta_{i,l+1}\delta_{j,l+1}-\delta_{i,l}\delta_{j,l}-\delta_{i,l+1}\delta_{j,l}\Big)
+2\Big(\delta_{j,l+1}-\delta_{j,l}\Big)\delta_{i> l}+\nn\\
+4\delta_{i,l}\delta_{j,l}-\delta_{i,j}\delta_{j,l+1}+5\delta_{i,l+1}\delta_{j,l}-2\delta_{i,l+1}\delta_{j,l+1}\nn\\
v^{ij}_l=2(N-l)\Big(\delta_{i,l}\delta_{j,l}-\delta_{i,l+1}\delta_{j,l+1}\Big)+
\Big(\delta_{j,l}-\delta_{j,l+1}\Big)\delta_{i> l}+\Big(\delta_{i,l}
-\delta_{i,l+1}\Big)\delta_{j> l}+6\delta_{i,l+1}\delta_{j,l+1}\nn\\
w^{ij}=2\delta_{ij}+2(N-j)\delta_{i,j-1}+2(i+2-N)\delta_{i,j+1} \label{zm}
\ee
With the help of this identity,
{\footnotesize
\be
 \! \! \! \! \! \!  \! \! \! \! \!
\hat L_{2} {\cal Z} =
\prod_{i=1}^N  \int
e^{-a_i^2/2} da_i \prod_{j=1}^{N-1} \int_0^\infty \! \! e^{-b_j} b_j^{\beta j-1} db_j
\left( \Big(2N\beta+3(1-\beta)\Big)\Phi_2+\beta\Phi_1^2
-  \tr \Phi^4 + \sum_{k=1} kt_k \,\tr \Phi^{k+2}\right)
\exp\left\{\sum_k t_k \,\tr \Phi^k\right\} =  \ \ \ \nn
\ee
\be
= \prod_{i=1}^N  \int
e^{-a_i^2/2} da_i \prod_{j=1}^{N-1} \int_0^\infty e^{-b_j} b_j^{\beta j-1} db_j
\left( \Big(2N\beta +3(1-\beta)\Big)\Big(\sum_{i=1}^N a_i^2 +2\sum_{j=1}^{N-1} b_j\Big)
+\beta\Big(\sum_{i=1}^N a_i\Big)^2
- \right.\nn \\ \left.
-  \sum_{i=1}^Na_i^4
-4\sum_{j=1}^{N-1} b_j (a_j^2+a_ja_{j+1}+a_{j+1}^2)
- 2\sum_{j=1}^{N-1} b_j^2 - 4\sum_{j=1}^{N-2} b_jb_{j+1}
+ \right.\nn \\ \left.
+  \sum_{i=1}^N a_i^3\frac{\p}{\p a_i}
+ \sum_{l=1}^{N-1} b_l \sum_{i,j=1}^N u_l^{ij}a_i \frac{\p}{\p a_j}
+\sum_{l=1}^{N-1} b_l \sum_{i,j=1}^N v_l^{ij}a_ia_j \frac{\p}{\p b_l}
+ \sum_{i,j=1}^{N-1} w^{ij}b_ib_j\frac{\p}{\p b_j}
\right)
\exp\left\{\sum_k t_k \,\tr \Phi^k\right\}
\nn
\ee
}

\bigskip

\noindent
Integration by parts gives:
{\footnotesize
\be
\hat L_{2} {\cal Z} = \prod_{i=1}^N  \int
e^{-a_i^2/2} da_i \prod_{j=1}^{N-1} \int_0^\infty e^{-b_j} b_j^{\beta j-1} db_j
\left( \Big(2N\beta +3(1-\beta)\Big)\Big(\sum_{i=1}^N a_i^2 +2\sum_{j=1}^{N-1} b_j\Big)
+\beta\Big(\sum_{i=1}^N a_i\Big)^2
- \nn\right. \\ \left.
-  \sum_{i=1}^Na_i^4
-4\sum_{j=1}^{N-1} b_j (a_j^2+a_ja_{j+1}+a_{j+1}^2)
- 2\sum_{j=1}^{N-1} b_j^2 - 4\sum_{j=1}^{N-2} b_jb_{j+1}
+  \sum_{i=1}^N (a_i^4-3a_i^2)
+ \right.\nn \\ \left.
+ \sum_{l=1}^{N-1} b_l \sum_{i,j=1}^N u_l^{ij}\Big(a_i a_j-\delta_{ij}\Big)
+\sum_{l=1}^{N-1}   \sum_{i,j=1}^N v_l^{ij}a_ia_j \Big(b_l - l\beta \Big)
+ \sum_{i,j=1}^{N-1} w^{ij}\Big(b_ib_j-\delta_{ij}b_j  - j\beta b_i \Big)
\right)
\exp\left\{\sum_k t_k \,\tr \Phi^k\right\}
= \nn
\ee
\be
\!\!\!\!\!\!\!\!\!\!\!\!\!\!\!\!\!\!\!\!
= \prod_{i=1}^N  \int
e^{-a_i^2/2} da_i \prod_{j=1}^{N-1} \int_0^\infty e^{-b_j} b_j^{\beta j-1} db_j
\left(\beta (2N -3)\Big(\sum_{i=1}^N a_i^2 +2\sum_{j=1}^{N-1} b_j\Big)
+\beta\Big(\sum_{i=1}^N a_i\Big)^2
-\beta\sum_{l=1}^{N-1}   \sum_{i,j=1}^N lv_l^{ij}a_ia_j
-\beta \sum_{i,j=1}^{N-1} jw^{ij}  b_i
- \right.\nn
\ee
}
\vspace{-.3cm}
\be
\left.
-4\sum_{j=1}^{N-1} b_j (a_j^2+a_ja_{j+1}+a_{j+1}^2)
- 2\sum_{j=1}^{N-1} b_j^2 - 4\sum_{j=1}^{N-2} b_jb_{j+1}
+ 6\sum_{j=1}^{N-1} b_j
+ \right.\nn \\ \left.
+ \sum_{l=1}^{N-1} b_l \sum_{i,j=1}^N u_l^{ij}\Big(a_i a_j-\delta_{ij}\Big)
+\sum_{l=1}^{N-1}   \sum_{i,j=1}^N v_l^{ij}a_ia_j b_l
+ \sum_{i,j=1}^{N-1} w^{ij}\Big(b_ib_j-\delta_{ij}b_j \Big)
\right)
\exp\left\{\sum_k t_k \,\tr \Phi^k\right\}
\label{eq:l2-act}
\ee

\noindent
which is zero with the choice of zero modes (\ref{zm}).
\end{itemize}

Formulas (\ref{zm}) may look a bit terse,
so we also provide expressions for the tensors $u$, $v$ and $w$
for first few $N$ in the Appendix.

\section{Zero modes and the full expression for (\ref{2id})}

For purposes of the present paper, we need a special selection of zero modes,
and this what we made in the previous section to simplify the formulas.
But they are always present in the tridiagonal formalism.
The reason for emerging zero modes is simple:
when one tries to express the $N$ eigenvalues through the $2N-1$ quantities $a_i,b_j$, one
inevitably gets ambiguities related to the deformations in the ``angular'' directions,
the zero modes. These zero modes manifest themselves in existence of the first order differential operators
that annihilate \textit{every} $\text{tr} \Phi^k$.
There are $2N-1-N=N-1$ such operators, but they differ by degree,
and only some of them matter for Virasoro constraint $L_n$,
which is of degree $n$ (we assume that $\deg a = 1$ and $\deg b = 2$).
As we saw in (\ref{rel-1}) and (\ref{rel0})
there are no annihilating operators in degrees -1 and 0.
There is one annihilating operator in degree 1,
\be\label{ao1}
\hat D_1:=\sum_{i=1}^{N-1}\hat \nabla_i,\ \ \ \ \ \ \
\hat \nabla_i:=b_i\left[(a_i-a_{i+1}){\p\over\p b_i}+{\p\over\p a_{i+1}}-{\p\over\p a_i}\right]
\ee
so that one can add to (\ref{zm1}) this operator with an arbitrary coefficient $c$.

In degree 2 one can add to (\ref{2id}) an arbitrary linear combination of $a_i\hat D_1$
and a new operator $\hat D_2$,
\be\label{ao2}
\hat D_2:=-\sum_{i=1}^{N-1}(a_i+a_{i+1})\hat \nabla_i+
\sum_{i=1}^{N-2}b_ib_{i+1}\left({\p\over\p b_i}-{\p\over\p b_{i+1}}\right)
\ee
Hence, there are $N+1$ zero modes
(generated by just two independent annihilation operators $\hat D_1$ and $\hat D_2$),
and, for possible future applications, we provide here the explicit expression of (\ref{2id}) with the linear combination of zero modes $\sum_i\xi_ia_i\hat D_1+\eta\hat D_2$ included:
\begin{align} \label{eq:full-k-2-operator}
  & \left(\sum_{i=1}^N a_i^3\frac{\p}{\p a_i}
  + \sum_{l=1}^{N-1} b_l \sum_{i,j=1}^N u_l^{ij}a_i \frac{\p}{\p a_j}
  +\sum_{l=1}^{N-1} b_l \sum_{j<i}^N v_l^{ij}a_ia_j \frac{\p}{\p b_l}
  + \sum_{i,j=1}^{N-1} w^{jk} b_ib_j\frac{\p}{\p b_j} \right)
  \tr \Phi^k = k\, \tr \Phi^{k+2} \\ \notag
  u^{ij}_k = &  \Big(2\delta_{i<k}+\xi_i+(2i-2+\eta)\delta_{i,j}\Big)
  (\delta_{k,j}-\delta_{k,j-1})  + (2 i - 5 + \eta)\Big( \delta_{k,j} \delta_{i,j+1}
  - \delta_{k,j-1} \delta_{i,k}\Big)+ 6 \delta_{k,j-1} \delta_{i,j} \\ \notag
  v^{ij}_k = & \Big(2+\xi_j+(2i-4+\eta)\delta_{i,j}\Big)(\delta_{k,i-1}-\delta_{k,i})+\xi_i(\delta_{k,j-1}-\delta_{k,j})+
  2\delta_{k,i}\delta_{j,i}
  \\ \notag
  w^{jk} = & 2 \delta_{i,j}
  + (-2 i + 6 - \eta) \delta_{i,j-1}
  + (2 i - 4 + \eta) \delta_{i,j+1}
\end{align}
This full formula is not needed for the derivation of Virasoro constraints,
still is an interesting identity of its own,
which deserves better understanding and generalization beyond the tridiagonal formalism.
The choice of zero modes (\ref{zm}) corresponds to $\eta=8-2N$, $\xi_i=-2$ in (\ref{eq:full-k-2-operator}).

Note that expressions (\ref{ao1})-(\ref{ao2}) provide us with only first two annihilating operators.
Moreover, at $N=2$ there is exactly one such operator $\hat D_1$, while $\hat D_2$ is linearly dependent since the second term in (\ref{ao2}) is absent in this case: $\hat D_2=-(a_1+a_2)\hat D_1$.
As we explained above, there are totally $N-1$ operators so that the remaining $N-3$ come at higher degrees.
Accordingly, the number of zero modes that should be tamed in the analysis of Virasoro constraints $L_n$
fastly grows with $n$.

\section{Non-equivalence of tridiagonal and diagonal measures}

One could  think that the measures in (\ref{triMAMO}) and (\ref{evMAMO}) are just equivalent
but they are not as we demonstrate in this section.
What happens is that they are the same when one integrates a peculiar function
which depend only on symmetric combinations of eigenvalues, i.e. on $\ \tr \Phi^k$.
In other words, the measures are equivalent {\it modulo} zero modes.

As everywhere in this paper, we concentrate on the case of real matrices,
i.e. symmetric and orthogonal rather than Hermitian and unitary,
see the reminder in sec.\ref{DEM}.
Also in this section, we use parametrization (\ref{Psi}) rather than (\ref{Phi}).
We restrict to representative  examples only, for an exhaustive
analysis, see the original papers  \cite{paper:DE-matrix-models-for-beta-ensembles}.

\subsection{Diagonal (eigenvalue) measure
\label{DEM}}

For $H=U^\dagger D U$,
the variation $\delta H = U^\dagger \delta D U + U^\dagger D\delta U - U^\dagger \delta U U^\dagger D U$
and
\be
\Tr \delta H^2 = \Tr \delta D^2 + 2\Tr \delta U U^\dagger \overbrace{[\delta D,D]}^{0}
 + 2 \Tr (\delta U U^\dagger D)^2 - 2\Tr (\delta U U^\dagger)^2 D^2
 = \nn \\
 = \sum_{i=1}^N \delta D_i^2 +   \sum_{i,j}^N (\lambda_i-\lambda_j)^2(\delta U U^\dagger D)_{ij}
\ee
so that
\be
\int F\{ p_k \} \  d^{N^2}\!H
= \int      F\left\{p_k\right\}\,
\Delta^2(\lambda)\prod_{i=1}^N d\lambda_i \int DU
\ee
where $p_k=\tr H^k = \sum_i \lambda_i^k$.
In the case of real matrices $H$, the $U$'s are orthogonal and
(the absolute value of) the
Vandermonde product
$\Delta(\lambda) = \prod_{i>j}^N (\lambda_i-\lambda_j)$
appears in the first power.

If one integrates the rotation invariant ($U$-independent) quantities,
which depend only on the eigenvalues $\lambda$ the integral over $U$
decouples, and can be ignored.
In any case, the measure splits into the $\lambda$-dependent and $U$-dependent parts.

\subsection{The case of $N=2$}

For tridiagonal measure the splitting is more tricky.

At $N=2$, the tridiagonal matrix is just a real-valued symmetric matrix,
$H = \left(\begin{array}{cc}a_1 & \b_1 \\ \b_1 & a_2 \end{array}\right)$.
It can be diagonalized with the help of rotation by one auxiliary angle
\begin{align}
  \left(\begin{array}{cc}a_1 & \b_1 \\ \b_1 & a_2 \end{array}\right)
  =
  \left(\begin{array}{cc} \cos\theta & \sin\theta \\ -\sin\theta & \cos\theta \end{array}\right)
  \left(\begin{array}{cc}\lambda_1 & 0 \\ 0 & \lambda_2 \end{array}\right)
  \left(\begin{array}{cc} \cos\theta & -\sin\theta \\ \sin\theta & \cos\theta \end{array}\right)
  = \left(\begin{array}{cc} \lambda_1 c^2 + \lambda_2 s^2 & (-\lambda_1 + \lambda_2) c s \\
    (-\lambda_1 + \lambda_2) c s & \lambda_1 s^2 + \lambda_2 c^2 \end{array}\right)
\end{align}
where $c=\cos\theta$ and $s =\sin\theta$.

Derivative of the map $(a_1, a_2, \b_1) \rightarrow (\lambda_1, \lambda_2, \theta)$ is
\begin{align}
  \left( \begin{array}{c} d a_1 \\ d a_2 \\ d \b_1 \end{array} \right)
    =
    \left( \begin{array}{ccc}
      c^2 & s^2 & 2 \b_1 \\
      s^2 & c^2 & - 2 \b_1 \\
      - c s & c s & \b_1 \frac{(c^2 - s^2)}{cs}
    \end{array} \right)
    \left( \begin{array}{c} d \lambda_1 \\ d \lambda_2 \\ d \theta \end{array} \right)
\label{N2da}
\end{align}
its Jacobian is equal to $-(\lambda_1 - \lambda_2)$, and one has
\begin{align}
  d a_1 da_2 \b_1^{2 \beta - 1} d \b_1 = -(-cs)^{2\beta}d\theta \cdot (\lambda_1 - \lambda_2)^{2 \beta} d\lambda_1 d\lambda_2
\end{align}
where we substituted $\b_1 =  -(\lambda_1 - \lambda_2) c s$.
Ignoring the angular part, one gets just $(\lambda_1 - \lambda_2)^{2 \beta} d\lambda_1 d\lambda_2$
in full accordance with (\ref{triMAMO1}).

The apparent $\theta$-independence of the quantities $\tr H^k$ is equivalent to the zero mode consideration
in the previous section.
Indeed, from (\ref{N2da})
\be
\frac{\p}{\p \theta} = \frac{\p a_1}{\p \theta} \frac{\p}{\p a_1} +
\frac{\p a_2}{\p \theta} \frac{\p}{\p a_2}+\frac{\p \b_1}{\p \theta} \frac{\p}{\p \b_1}
= 2\b_1\left( \frac{\p}{\p a_1}-\frac{\p}{\p a_2}\right) -(a_1-a_2) \frac{\p}{\p \b_1}
= -\frac{2}{\b_1} \hat \nabla_1
\label{dtheta}
\ee
Note that the zero mode operator  $\hat \nabla_1$ in (\ref{ao1}) is defined in terms of another $b$-variable,
which is $ \b_1^2$ in the notation of this section:
\be
\hat \nabla_1= \b_1^2\left[(a_1-a_{2}){\p\over\p  \b_1^2}
+{\p\over\p a_{2}}-{\p\over\p a_1}\right]=-\frac{\b_1}{2  }{\p\over\p\theta}
\ee

\subsection{The case of $N=3$}

This is the crucial example.
In this case, we can diagonalize symmetric matrix
with the help of orthogonal one, $H=\Theta\cdot \Lambda\cdot \Theta^T$,
which we can parameterize, say, by the three Euler angles.
Then
\begin{align}
  \left(\begin{array}{ccc}a_1 & \b_1 & \b_{12} \\ \b_1 & a_2 & \b_2 \\ b_{12} & \b_2 & a_3 \end{array}\right)
  =
  \overbrace{\left(\begin{array}{ccc}
  \cos\theta_3 & \sin\theta_3 & 0 \\ -\sin\theta_3 & \cos\theta_3 & 0 \\ 0&0&1
  \end{array}\right)
  \left(\begin{array}{ccc}
  1&0&0 \\ 0& \cos\theta_2 & \sin\theta_2  \\ 0& -\sin\theta_2 & \cos\theta_2
  \end{array}\right)
  \left(\begin{array}{ccc}
  \cos\theta_1 & \sin\theta_1 & 0 \\ -\sin\theta_1 & \cos\theta_1 & 0 \\ 0&0&1
  \end{array}\right)}^{\Theta}
  \cdot \nn \\
  \cdot
  \left(\begin{array}{ccc}\lambda_1&0&0 \\ 0&\lambda_2&0 \\0&0&\lambda_3 \end{array}\right)
  \left(\begin{array}{ccc}
  \cos\theta_1 & -\sin\theta_1 & 0 \\ \sin\theta_1 & \cos\theta_1 & 0 \\ 0&0&1
  \end{array}\right)
  \left(\begin{array}{ccc}
  1&0&0 \\ 0& \cos\theta_2 & -\sin\theta_2  \\ 0& \sin\theta_2 & \cos\theta_2
  \end{array}\right)
  \left(\begin{array}{ccc}
  \cos\theta_3 & -\sin\theta_3 & 0 \\ \sin\theta_3 & \cos\theta_3 & 0 \\ 0&0&1
  \end{array}\right)
\end{align}

The Jacobian is again proportional to the Vandermonde determinant:
\be
da_1da_2da_3d\b_1d\b_2d\b_{12} =
\sin\theta_2 d\theta_1d\theta_2d\theta_3
\cdot (\lambda_2-\lambda_1)(\lambda_3-\lambda_1)(\lambda_3-\lambda_2)d\lambda_1d\lambda_2d\lambda_3
\label{Jac3}
\ee
However, expressions for $\b_1,\b_2,\b_{12}$ are now more complicated.
Still, one can notice that $\b_{12}$ does not depend on $\sin\theta_2$ but only on $\cos\theta_2$,
and $\b_{12}=0$ implies that
\be
c_2 = \frac{c_1s_1c_3(\lambda_1-\lambda_2)}{s_3(c_1^2\lambda_2+s_1^2\lambda_1-\lambda_3)}
\ee
Note a remarkable simplicity of this expression.
However, expressions for $\b_1$ and $\b_2$ are rather complicated even at this tridiagonal locus.
Still, somewhat miraculously, three particular combinations are spectacularly simple:
\be
\left.\b_1\b_2^2\right|_{\b_{12}=0}
= (\lambda_1-\lambda_2)(\lambda_1-\lambda_3)(\lambda_2-\lambda_3) \cdot
\Theta_{31}\Theta_{32}\Theta_{33}
\label{b1b22}
\ee
\be
\left.(a_1-a_3)\cdot \b_1\b_2\right|_{\b_{12}=0}
= -(\lambda_1-\lambda_2)(\lambda_1-\lambda_3)(\lambda_2-\lambda_3) \cdot
\Theta_{21}\Theta_{22}\Theta_{23}
\label{b1b2}
\ee
\be
\left.\b_1^2\b_2 \right|_{\b_{12}=0}
= -(\lambda_1-\lambda_2)(\lambda_1-\lambda_3)(\lambda_2-\lambda_3) \cdot
\Theta_{11}\Theta_{12}\Theta_{13}
\label{b12b2}
\ee
Also simple is the Jacobian similar to (\ref{Jac3}), but for 5 variables, i.e. taking into account the constraint $b_{12}=0$: let us define the Jacobian $J(a,b;\lambda,\Theta)$ as
\be
\prod_{i}da_i\prod_jd\b_j =J(a,\b;\lambda,\Theta)\prod_i d\lambda_id^2\Theta_{k}
\ee
where $d^2\Theta_k$ is the surface elements of the sphere $S^2$ in the space of three elements $(\Theta_{k,1},\Theta_{k,2},\Theta_{k,3})$ of {\it any} line of the rotation matrix, since $\sum_{i=1}^3\Theta_{k,i}=1$ for any orthogonal matrix. This Jacobian is equal to
\be
J(a,\b;\lambda,\Theta)={\b_1\b_2\over\Theta_{k,1}\Theta_{k,2}\Theta_{k,3}}
\ee

If something $\Theta$-independent is integrated,
like a function of $\Tr \Psi^k$,
from the first of these expressions one gets exactly (\ref{triMAMO1}).
In fact, such an integral can be written in two above forms, (\ref{b1b22}) and (\ref{b12b2}),
which give equal answers
because of the symmetry of the integrand w.r.t. the permutations of ${\rm b}_i$.
Angular averages of $\Theta$-products provide just overall constants,
which are in fact equal.

Another argument involves the zero modes.
Generalizing (\ref{dtheta}), the two annihilating operators (\ref{ao1})-(\ref{ao2})
are again made from the angular direction operators ${\p\over\p\theta_1}$, ${\p\over\p\theta_3}$:
\be
\hat D_1={1\over 2}\left({s_1^2\over 2}{\p a_2\over\p \theta_1}-\b_1c_1^2\right){\p\over\p\theta_1}+{s_3c_3(\lambda_1-\lambda_2)\over 2}{\p
\over\p\theta_3}\\
\hat D_2=\xi\cdot{\p\over\p\theta_1}+{s_3c_3(\lambda_1^2-\lambda_2^2)\over 2}{\p
\over\p\theta_3}
\ee
and, hence give rise to zero modes when acting on invariant expressions depending on $\Tr \Psi^k$ only. Here the coefficient $\xi$ is quite involved:
\be
\xi:={c_1^2\over 2s_1^2}\left(2\lambda_3-(s_1^2+1)(\lambda_1+\lambda_2)\right)\b_1-{c_1\over 2s_1}\left({\b_1^2\over s_1^2}+
(\lambda_1-\lambda_3)(\lambda_2-\lambda_3)\right)+{s_2^2\over 4}(\lambda_1+\lambda_2){\p a_2\over\p \theta_1}
\ee

\subsection{Generic $N$}

Let us diagonalize the tridiagonal matrix $\Psi$, (\ref{Psi}):
\be
\Psi=\Theta\cdot \Lambda\cdot \Theta^T
\ee
where $\Theta$ is an orthogonal matrix, and $\Lambda$ is a diagonal matrix with $\lambda_i$ on the diagonal. This can be done since the matrix $\Psi$ is symmetric.
Denote matrix elements (angles) from the {\it last} row of $\Theta$ as $\Theta_i$.
Then, the following generalizations of (\ref{b1b22}) and (\ref{Jac3}) are correct \cite{paper:DE-matrix-models-for-beta-ensembles}:
\begin{itemize}
\item The Vandermonde determinant of $\lambda_i$ can be rewritten in terms of $\b_i$ and $\theta_i$ in a simple way:
\be
\Delta(\lambda)={\prod_{i=1}^{N-1}\b_i^i\over\prod_{i=1}^N\Theta_i}
\ee
\item The Jacobian $J(a,\b;\lambda,\Theta)$ of change of variables from $(a_i,\b_i)$ to $(\lambda_i,\Theta_i)$ is given by a similar ratio:
\be
J(a,\b;\lambda,\theta)={\prod_{i=1}^{N-1}\b_i\over\prod_{i=1}^N\Theta_i}
\ee
\end{itemize}
From these two identities, one immediately obtains
\be
\prod_{i=1}^Nda_i\prod_{i=1}^{N-1}\b_i^{2\beta i-1}d\b_i=
\left(\prod_{i=1}^N\theta_i^{2\beta-1}d\Theta_i\right)\Delta(\lambda)^{2\beta}\prod_{i=1}^Nd\lambda_i
\ee
The factor in the brackets at the r.h.s. is  an angular factor which contributes
just an irrelevant constant when one integrates the invariant functions,
depending only on the quantities $\tr \Psi^k$,
which are annihilated by the zero mode generators.

{\subsection{A puzzling anomaly} \label{sec:anomaly}

The mystery of tridiagonal representation is not exhausted by zero modes.
An apparent puzzle is the appearance of $b_j^{\beta j}$ in the measure (\ref{triMAMO}),
which explicitly breaks the invariance $j\longleftrightarrow N-j$.
This invariance obviously forbids exactly the {\it linear} dependence on $j$ in the exponent
but mysteriously it is exactly the one which appears.
The symmetry is restored by existence of a complementary formulas, like (\ref{b1b22}) and (\ref{b12b2}),
but it is  ``spontaneously broken" in tridiagonal averages.
This {\it anomaly} is one of the first issues to address in the future study of the tridiagonal formalism.

\section{On potential use of the tridiagonal approach}

We emphasize once again that our
derivation based on the Virasoro constraints is straightforwardly
generalized to arbitrary non-Gaussian potentials
generated by the angular independent quantities
$\tr \Phi^k = \sum_{i=1}^N\lambda_i^k$,
where $\Phi$ is given by \eqref{Phi}.
The averages in the $\beta$-ensemble defined as
\be
\Big< F\{p_k\}\Big> _{T_k}:=  \int_{-\infty}^\infty
\Delta(\lambda)^{2\beta}\,  \prod_{i=1}^N  e^{-\sum_{k=1}^d T_k\lambda_i^k} d\lambda_i
\cdot F\{p_k\}
\label{evNG}
\ee
can be calculated as
\be
\langle F\{p_k\} \rangle  \sim
\int_{{\cal C}^\prime } \prod_{i=1}^N  d a_i
\prod_{j=1}^{N-1}
 b_j^{\beta j - 1} db_j
\cdot e^{-\sum_{k=1}^d T_k \tr \Phi^k} F\{\tr \Phi^k\},
\label{triNG}
\ee
Though somewhat unusual and technically frightening,
this tridiagonal description of the $\beta$-ensemble
has a crucial advantage over the eigenvalue formula (\ref{evNG}).
Namely, with the help of explicit integrals like
\be
\int_0^\infty e^{-b} b^{\beta j + n}db = \Gamma\left(1 + \beta j + n\right)
\ee
it can lead to expressions that are \textit{symbolic} in $\beta$,
while with \eqref{evNG} one usually  performs calculations for a particular integer $\beta$
and then analytically continues the result.
{\bf The tridiagonal realization} allows one to put the analytical continuation under control,
and {\bf directly provides formulas for a generic non-integer $\beta$}.

Still there is a subtlety.
The non-perturbative proof of sec.~\ref{sec:non-perturbative-derivation}
is literally carried over to the non-Gaussian case {\it provided} one chooses sets of integration
contours ${\cal C}$ and ${\cal C}^\prime$ properly. The crucial element of the proof,
the integration by parts without appearance of any boundary terms is valid whenever
the contours ${\cal C}$ and ${\cal C}^\prime$ end at zeroes of the respective
(non-Gaussian) integration measures.
At the loci $b_j = 0$, the measure vanishes provided $\beta > 1$.
However, for $\beta < 1$ more subtle analysis is required, and new, previously unseen, exotic
phase of the $\beta$-ensemble may appear.
The exploration of this intriguing direction is left
for future research.

\section{Conclusion}

In this paper, we suggested to derive the tridiagonal realization of $\beta$-ensembles
due to I.Dumitriu and A.Edelman \cite{
  paper:DE-matrix-models-for-beta-ensembles,
  paper:K-on-gaussian-random-matrices-coupled-to-the-discrete-laplacian
},
from the superintegrability property:
that the Jack averages are expressed through the same Jack polynomials.
This can help to demystify the tridiagonal formalism, and open a route
to many fruitful applications.
We also suggested a proof of this realization with the help of Virasoro constraints,
whose logic may be more compelling for people with field theoretic background.
Further generalization to the Macdonald functions \cite{
  book:M-symmetric-functions-and-hall-polynomials
} and (generalized) Shiraishi functions \cite{
  paper:S-affine-screening-operators,
  paper:AKMM-shiraishi-functor-and-non-kerov
  }
can be one of the next steps to perform. For other interesting aspects of tridiagonal representations, see
\cite{leonard1982orthogonal,paper:terwilliger2001two
}
and references therein.

\section*{Acknowledgements}

This work was supported by the Russian Science Foundation (Grant No.20-12-00195).

\section*{Appendix}

In this Appendix, we provide explicit examples for the first few $N$
of the tensors $u$, $v$ and $w$
given by the general but rather complicated
formulas \eqref{zm}.
These are the answers with the zero modes adjusted to produce the $L_2$ constraint.
From these examples, one can see how the tridiagonality reveals
itself in a peculiar way in this Virasoro operator.
\be
N=2:
\nn
\ee
\be
u_1 
=\left(\begin{array}{cc} 2 & 1 \\ 1 & 2 \end{array}\right), \ \ \
 v_1 
= \left(\begin{array}{cc} 2 & 1 \\ 1 & 2 \end{array}\right), \ \ \ w=2  \
\ee

\be
N=3:  \ \ \ \ \ \ \ \
\nn
\ee
\be
u_1 
 = \left(\begin{array}{ccc}  0 & 3 & 0 \\ -1 & 4 & 0 \\ -2 & 2 & 0 \end{array}\right),
 \ \ \ \ \ \
u_2 
 = \left(\begin{array}{ccc}  0 & 0 & 0 \\ 0 & 2 & 1 \\ 0 & 1 & 2 \end{array}\right),
 \ \ \ \ \ \
\nn  \ \ \ \ \ \
\!\!\!\! v_1 
 = \left(\begin{array}{ccc}  4 & 1 & 1 \\ 1 & 0 & -1 \\ 1 & -1 & 0 \end{array}\right),\ \ \
v_2 
 = \left(\begin{array}{ccc}  0 & 0 & 0 \\ 0 & 2 & 1 \\ 0 & 1 & 2 \end{array}\right),
\nn
\ee
\be
w
= \left(\begin{array}{cc}   2 & 2 \\ 2 & 2 \end{array}\right)
\ee

\be
N=4:  \ \ \ \ \ \ \ \
\nn
\ee
\be
u_1 = \left(\begin{array}{cccc}
-2&5&0&0\\ -3&6&0&0\\-2&2&0&0\\-2&2&0&0
\end{array}\right), \ \ \ \
u_2 = \left(\begin{array}{cccc}
0&0&0&0\\0&0&3&0\\0&-1&4&0\\0&-2&2&0
\end{array}\right), \ \ \ \
u_3 = \left(\begin{array}{cccc}
0&0&0&0\\0&0&0&0\\0&0&2&1\\0&0&1&2
\end{array}\right),
\ee
\be
v_1 = \left(\begin{array}{cccc}
6&1&1&1\\1&-2&-1&-1\\1&-1&0&0\\1&-1&0&0
\end{array}\right), \ \ \ \
v_2 = \left(\begin{array}{cccc}
0&0&0&0\\0&4&1&1\\0&1&0&-1\\0&1&-1&0
\end{array}\right), \ \ \ \
v_3 = \left(\begin{array}{cccc}
0&0&0&0\\0&0&0&0\\0&0&2&1\\0&0&1&2
\end{array}\right), \ \ \ \ \ \ \ \ \ \
w = \left(\begin{array}{cccc}
2&4&0\\0&2&2\\0&2&2
\end{array}\right)
\nn
\ee
Clearly when $N$   increases by one the only new matrices are
$u_1,v_1$ and the first line/row in $w$.

Coming to explicit expressions,
especially simple is the description of symmetric matrices $v$.
The ``new" $v_1$ is defined from the condition that the sum $\sum_{l=1}^{N-1} v_l$
is a diagonal matrix with the entries $\ {\rm diag}(2N-2,2,2,\ldots, 2)\ $
and unities in the first row and column, i.e.
\be
v_1 = \left(\begin{array}{ccccccc}
2N-2&1&1&\ldots& 1 \\ 1&2&0&\ldots&0 \\1&0&2&\ldots&0 \\ &&\ldots \\1&0&0&\ldots& 2
\end{array}\right)\  - \ \sum_{l=2}^{N-1}v_l
= \left(\begin{array}{ccccccc}
2N-2&1&1&1&\ldots& 1 \\ 0&6-2N&-1&-1&\ldots&-1 \\1&-1&0&0&\ldots&0 \\1&-1&0&0&\ldots&0
\\ &&\ldots \\1&-1&0&0&\ldots& 0
\end{array}\right)
\ee
where $v_l$ with $l=2,\ldots,N-1$ are inherited from the previous value of $N$
by adding the first zero line and the first zero column.

Similarly inherited are $u_l$ with $l=2,\ldots,N-1$,
while $u_1$ has non-vanishing entries in the first two columns,
all lines are just $(-2,2,0,0,\ldots,0)$ except for the first two, which are
$(6-2N, 2N-3, 0 ,0, \ldots, 0)$ and $(5-2N, 2N-2, 0 ,0, \ldots, 0)$, i.e.
\be
u_1 = \left(\begin{array}{ccccccc}
6-2N&2N-3&0&\ldots& 0 \\ 5-2N&2N-2&0&\ldots&0 \\-2&2&0&\ldots&0 \\ &&\ldots \\-2&2&0&\ldots& 0
\end{array}\right)
\ee

The $w$ matrix has the most sophisticated structure. Namely, for $N=5$  and $N=6$ it is
\be
w_{N=5} =  \left(\begin{array}{cccc}
{\bf 2}&6&0& 0 \\ -2&{\bf 2}&4&0 \\0&0&{\bf 2}&2 \\ 0&0&2&{\bf 2}
\end{array}\right), \ \ \ \ \ \ \
w_{N=6} =  \left(\begin{array}{ccccc}
{\bf 2}&8&0&0& 0 \\ -4&{\bf 2}&6&0&0 \\0&-2&{\bf 2}&4&0 \\0&0&0&{\bf 2}&2 \\ 0&0&0&2&{\bf 2}
\end{array}\right)
\ee
which makes transparent its tridiagonal structure
with $2$ standing at the diagonal (boldfaced), and $6+2i-2N$, $2N-2i$, at the two adjacent sub-diagonals:
\be
w = \left(\begin{array}{cccccccccc}
{\bf 2}& 2N-4 & 0 & 0& \ldots &0&0&0&0&0 \\
8-2N& {\bf 2}&2N-6& 0 & \ldots &0&0&0&0&0 \\
0&10-2N&{\bf 2}&2N-8 & \ldots &0&0&0&0&0 \\
0 & 0& 12-2N & {\bf 2} &\ldots &0&0&0&0&0 \\ \\ \\
&&\ldots \\ \\ \\
0&0&0&0& \ldots & -4&{\bf 2}&6&0&0 \\
0&0&0&0& \ldots & 0&-2&{\bf 2}&4&0 \\
0&0&0&0& \ldots & 0&0&0&{\bf 2}&2 \\
0&0&0&0& \ldots & 0&0 & 0 & 2&{\bf 2}
\end{array}\right)
\ee

This completes the description of the particular choice of zero modes (\ref{zm}) for the $L_2$-constraint.



\end{document}